\documentclass[draftpaper]{bio}

\volume{?}
\issue{?}
\jno{bio000}
\pyear{2012}

\begin{document}
\title{Optimal sampling ratios in comparative diagnostic trials}
 %
\author{Ting Dong}
\address{Department of Statistics, }
\address{George Mason University,
Fairfax, VA 22030}

\author{Liansheng Tang $^\ast$}
\address{Department of Statistics, }
\address{George Mason University,
Fairfax, VA 22030}
\email{ltang1@gmu.edu}

\author{William F. Rosenberger}
\address{Department of Statistics, }
\address{George Mason University,
Fairfax, VA 22030}

\raggedbottom

\maketitle
\newpage
\begin{abstract}
{Diagnostic trials evaluating a single marker or comparing two markers often employ an arbitrary sampling ratio between the case and the control groups. Such a ratio is not always an efficient choice when the goal is   to  maximize the power or to minimize the total required sample size.  Instead, optimal sampling ratios, discussed by \cite{Janes:06},  offer  a better alternative for one-marker trials.    In this paper we focus  on  comparative diagnostic trials  which are frequently employed to compare    two  markers with continuous or ordinal results.  We derive  explicit expressions for the optimal sampling ratio based on a common variance structure shared by many existing summary statistics of the receiver operating characteristic  (ROC) curve.    Estimating the optimal ratio   requires  either pilot data   or  parametric model assumptions; however,  pilot data are often unavailable at the planning stage of diagnostic trials. In the absence of pilot data, some
distributions have to be assumed for carrying out the calculation.
An optimal ratio from an incorrect distributional assumption may lead
to an underpowered study. We propose   a two-stage procedure to   adaptively estimate  the optimal ratio in comparative diagnostic trials without   pilot data or  assuming parametric distributions.   We illustrate the properties of the proposed method   through theoretical proofs and extensive simulation studies. We  use an example in cancer diagnostic studies to illustrate the application of our method. We find that our method increases the power, or reduces the required overall sample size dramatically.}
\end{abstract}
AUC; Diagnostic accuracy; Internal pilot data; Two-stage design

\section{Introduction}


Diagnostic trials  estimate the diagnostic
accuracy of a  marker or  compare the diagnostic accuracy of two markers. For example, in a diagnostic trial by \cite{Hendrick:08},
investigators compared  the accuracy of digital mammography with
screen-film mammography. \cite{Pepe:01} refer to
these trials as phrase III diagnostic trials. In these
trials,   the true disease status of subjects is known. To evaluate the diagnostic  accuracy of a binary marker, sensitivity and specificity  are   used. Sensitivity is the probability
of having a positive test result for a case subject.
Specificity is the probability of having a negative test result for a control subject. The false positive rate (FPR) is
$1-$specificity. For continuous markers, we obtain sensitivity
and false positive rate (FPR) based on a threshold that distinguishes the test result
as being positive or negative. A varying
threshold allows a number of  sensitivities and FPRs to be computed
simultaneously. The receiver operating characteristic (ROC) curve is
a plot of sensitivity versus FPR for all possible thresholds.

Typically the ratio between the number of cases versus the number of controls is fixed in advance. Most diagnostic trials apply an equal case-control ratio; for example,  a lung cancer prevention trial recruited  71 prostate cancer cases and 71 age-matched controls without  cancer \citep{Etzioni:03}. A diagnostic study in \cite{Hendrick:08} compared the accuracy of digital mammography with  screen-film mammography using  equal numbers of breast cancer patients and controls.   In a colorectal cancer-screening study,  about the same number of  colorectal cancer patients and non-cancer subjects were used to identify   markers \citep{Janes:05}. The equal ratio, however, may not be optimal in maximizing the test power or minimizing the total required sample size.
 A  procedure
proposed by \cite{Janes:06} estimates the optimal ratio for evaluating a continuous
marker. The ratio is optimal with regard to
minimizing the variance, or maximizing the
power for a fixed total required sample size.  Equivalently, the optimal ratio  minimizes the total required sample size with a fixed power.   To the best
of our knowledge, their method is the first attempt to identify the
optimal sampling ratio in diagnostic trials. However,
since the optimal ratio is derived using the first derivative of the
ROC curve, their method cannot  be used for  ordinal data which often
occur in  medical imaging  studies. More importantly,
pilot data are required to estimate the optimal ratio.  In the absence of pilot data, some
distributions have to be assumed for carrying out the calculation.
An optimal ratio from an incorrect distributional assumption may lead
to an underpowered study. In addition, optimal ratios for comparative diagnostic trials are of interest to investigators, but have not been discussed in the literature.

In this paper we derive the optimal sampling ratio of cases to controls in comparative diagnostic trials. The proposed optimal ratio is based on a common variance structure shared among existing  ROC summary statistics. Special cases of these statistics include the nonparametric area under the ROC curve (AUC) statistic  proposed by   \cite{Delong:88} and the weighted AUC statistic  by \cite{wgjj:89}. These statistics have been applied in the sequential diagnostic trial design by \cite{liu:03} and \cite{liu:08}. The calculation of the optimal sampling ratio  requires either parametric model assumptions or pilot data. When the parametric model is incorrectly specified, the resulting ratio may not give the optimal power or the minimal required sample size. It is desirable to re-calculate the optimal ratio when data become available during the trial. We propose a two-stage method to incorporate  the idea of internal pilot data, reviewed in \cite{proschan:04}. We assume  a parametric model at the beginning of the trial to obtain the initial optimal ratio. This ratio is used to sample the cases and controls at the first stage. When sufficient observations are available, the optimal ratio is re-calculated at the second stage, and the numbers of cases and controls are adjusted accordingly.
We show that although the optimal ratio is updated during a diagnostic trial, the analysis at the end of the trial can be carried out in the same fashion as in the traditional trial without affecting  the nominal  type I error rate.

 The  paper is organized as follows. In Section \ref{sec:ratio}, we start with   the optimal ratio for comparative diagnostic trials based on   common ROC  statistics. We then present the explicit expressions of the optimal ratios for comparing  AUCs and for comparing weighted AUCs. In Section \ref{sec:2stage}, we propose a two-stage procedure to adaptively estimate the optimal sampling ratio using the internal pilot data. We illustrate  the power increase and the savings on the overall required sample size using  the proposed method through a cancer example in Section \ref{sec:ex}.  Section \ref{sec:sim} investigates  the small sample performance of the proposed procedure in maintaining the  nominal type I error rate and increasing the power.  Some discussion is presented in Section \ref{sec:dis}.
\section{Optimal sampling  ratio}\label{sec:ratio}
Suppose we have $N$ subjects with $m$ cases and $n$ controls. Each subject is measured by diagnostic test $\ell$ ($\ell=1,2$). We define the $i$th case as $X_{\ell i}$, where $i=1, \ldots, m$, and the $j$th control as
$Y_{\ell j}$, where $j=1, \ldots, n$. The joint cumulative survival functions for cases are $(X_{1i}, X_{2i}) \sim S_{d}(x_1, x_2)$  and the joint cumulative survival functions for controls are $(Y_{1j}, Y_{2j}) \sim S_{\bar{d}}(y_1, y_2)$. Their marginal survival distributions are $X_{\ell i} \sim S_{d,\ell}(x)$  and $Y_{\ell j} \sim S_{\bar{d}, \ell}(y)$  respectively. For the threshold $c$ varying in $(-\infty, +\infty)$, the sensitivity is $ S_{d, \ell}(c)=Pr(X_{\ell i} > c) $,  and  the FPR is
$
S_{\bar{d}, \ell}(c)= Pr(Y_{\ell j} > c ).$  Subsequently,  the ROC curve for test $\ell$ is defined as
 $ R_\ell(u) = S_{d, \ell}(S_{\bar{d}, \ell}^{-1}(u))$, where the FPR, $u$, falls within $[0,1]$.

 Summary measures for a single ROC curve include the area under the ROC curve (AUC), the partial AUC (pAUC), and the weighted AUC (wAUC).  The AUC gives the probability that a measurement randomly selected from the case group is greater than the measurement randomly selected from the control group \citep{Bamber:75,hanley:82}; that is,
 $Pr(X>Y)=\int_0^1S_{d}\left\{S_{\bar{d}}^{-1}(u)\right\}du.$
 The wAUC by \cite{wgjj:89} is given by
  \begin{equation}\label{eq:weighted}
  \Omega=\int_0^1S_{d}\left\{S_{\bar{d}}^{-1}(u)\right\}dW(u),
  \end{equation}   where $W(u)$ is a probability measure.  We  let $W(u)$   be a point $u_0$, a FPR, to calculate the sensitivity of a test, or $ W(u)=u$, where $u \in (0,1)$, to estimate the AUC. When $ W(u)=(u-u_0)/(u_1-u_0)$, where $u \in (u_0, u_1)$, (\ref{eq:weighted}) gives the partial AUC.

The   statistics for comparing markers might be parametric, e.g., the binormal model of
\cite{dorfman:69}, semiparametric \citep{zou:97, tang2:08}, or
nonparametric  \citep{liu:03, Delong:88, hanley:83, wgjj:89}.
Let $\theta$ be the  parameter in
the ROC comparison, and $\hat \theta$ be the estimator. Based on the variance expressions for these ROC statistics,   we identify the following common structure for the variance of all these ROC statistics when the sample sizes get large:
\begin{equation}\label{eq:gvar}
var(\hat \theta)= \frac{v_{x}}{m}+\frac{v_{y}}{n},
\end{equation} where $v_{x}$ is the variance associated with  measurements of case patients and $v_y$ is  the variance related to control patients. In this paper we use the nonparametric statistics by  \cite{Delong:88} and \cite{wgjj:89}. We  present the variance expressions for these statistics    in Section \ref{sec:ratio1} and \ref{sec:ratio2}. One may refer to other aforementioned articles for the same variance structure  of parametric and semiparametric ROC statistics.

Given  the variance structure in (\ref{eq:gvar}),   the total required sample size in a diagnostic trial can be minimized using an optimal sampling ratio when the variance is fixed. In other words, the power for comparing two markers can be maximized using this optimal sampling ratio.
Suppose the total required sample size in the diagnostic trials is   $N=m+n$, the sampling ratio is $r=m/n$. Let the  variance of $\hat{\theta}$ is a fixed constant, $a$. Since
$ m=rn= Nr/(1+r),$  it follows that
$$v_{x}/m+v_{y}/n
= \frac{1+r}{N}(v_{x}/r+v_{y})=a.$$ The total required sample size can then be expressed as
$$N=\frac{1+r}{a}(v_{x}/r+v_{y}).$$
To minimize $N$, we take first derivative with respect to $r$ and equate it to zero. We obtain the following equation:
\begin{eqnarray}
v_{y}/a-v_{x}/ar^{-2}=0.\nonumber
\end{eqnarray}
By solving the equation above, the optimal  sampling ratio is obtained as
\begin{equation}\label{eq:opti}
r^*=\sqrt{\frac{v_x}{v_y}}.
\end{equation}
The optimal sampling  ratio is analogous to
the Neyman   allocation ratio  for clinical
trials which has been widely used to save the overall sample size for a fixed power. Interested readers can refer to  \cite{jennison:00} and \cite{Rosenberger:02}.
\subsection{Optimal sampling  ratio for   comparing two continuous markers}\label{sec:ratio1}
The difference between two wAUCs, $ \Delta= \Omega_1- \Omega_2$, is used in \cite{wgjj:89} to compare the wAUCs for  continuous data. Here the estimator $\hat\Omega_\ell$ of $\Omega_\ell,$ for $ \ell=1,2,$ is obtained by substituting the empirical function estimators in  (\ref{eq:weighted}). The resulting $\Delta$-statistic is given by $\hat\Delta=\hat\Omega_1-\hat\Omega_2$. Let $w_i$ be $\int_{0}^{1}[S_{d,1}(S_{\bar{d},
1}^{-1}(u)) - I(X_{1i} \leq  S_{\bar{d}, 1}^{-1}(u))-
S_{d,2}(S_{\bar{d}, 2}^{-1}(u)) + I(X_{2i} \leq  S_{\bar{d}, 2}^{-1}(u))]dW(u)$, and let $v_j$ be $\int_{0}^{1}
\{R'_{1}(u)[I(Y_{1j} \leq  S_{\bar{d}, 1}^{-1}(u)) - u] -  R'_{2}(u)[I(Y_{2j} \leq  S_{\bar{d}, 2}^{-1}(u)) - \!u]\}dW(u)$,
 \cite{Tang:08} further study the $\Delta$-statistic and show that for large sample sizes,
  $\hat\Delta $ is
asymptotically equivalent to
\begin{equation}\label{eq:wivj}
\frac{1}{m}\sum^{m}_{i=1}w_{i} +\frac{1}{n}\sum^{n}_{j=1}v_{j}+(\Omega_{1}-\Omega_{2}).
\end{equation}
Since $w_{i}$'s are $i.i.d.$ random variables corresponding
to measurements of case patients and $v_{j}$'s are also $i.i.d.$
random variables related to measurements of control subjects, (\ref{eq:opti}) gives the optimal ratio for comparing the difference between wAUCs:
\begin{equation}\label{eq:rwauc}
r^*=\sqrt{\frac{var(w_{i})}{var(v_{j})}},
\end{equation}
where $var(w_{i})$ is given by the following expression:
\begin{eqnarray}\label{eq:vx}
var(w_{i})&=& \sum_{\ell =1}^2\left ( \int_0^1\!\!\!\int_0^1 S_{d,\ell}
\{S_{\bar{d}, \ell}^{-1}(s\wedge t)\}dW(s)dW(t) - \left [\int_0^1
 S_{d,\ell} \{S_{\bar{d}, \ell}^{-1}(s )\}
  dW(s)\right ]^2 \right )\nonumber\\
 && -2 \!\int_0^1\!\!\!\int_0^1 \!\!  \left[S_{d}\{S_{\bar{d}, 1}^{-1}(s), S_{\bar{d}, 2}^{-1}(t)
\}\!-\!S_{d, 1}  \{S^{-1}_{\bar{d}, 1}(s)\}S_{d, 2}
 \{S_{\bar{d}, 2}^{-1}(t)\}\!\right]\!dW(s)\!dW(t),\nonumber
\end{eqnarray}and $var(v_{j})$ is  given by the following expression:
\begin{eqnarray}\label{eq:vy}
var(v_{j})&=&  \sum_{\ell =1}^2 \left[\int_0^1\!\!\!\int_0^1
 R'_\ell(s)R'_\ell(t) (s\wedge
t)dW(s)dW(t) -\left \{\int_0^1r_\ell(s)s dW(s) \right \}^2\right]\nonumber\\
&& -2 \int_0^1\!\!\!\int_0^1
 R'_1(s)R'_2(t)[S_{\bar{d}}\{S^{-1}_{\bar{d}, 1}(s), S_{\bar{d}, 2}^{-1}(t)
\}-st ]dW(s)dW(t),\nonumber
\end{eqnarray} with the derivative of $ROC_\ell(u)$, $
R'_\ell (u)=    {S'_{d, \ell}} \{ {S^{-1}_{\bar{d}, \ell}} (u)\}/
{S'_{\bar{d}, \ell}} \{ {S ^{-1}_{\bar{d}, \ell}}(u)\}$.

Since $\hat\Delta$ compares AUCs, partial AUCs or sensitivities  at a particular FPR, we discuss  the optimal ratios for these special cases   by specifying  corresponding weight functions.  When we let the weight function be $W(u)=u$, for $0<u<1$, $\hat\Delta$ compares the AUCs. The optimal ratio in (\ref{eq:rwauc}) implies that the following ratio between the case and the control maximizes the power for comparing the AUCs:
 $$r_{A}^{*}=\sqrt{\frac{v_x^A}{v_y^A}}, $$ where    $v_x^A$ and $v_y^A$ have the following expressions as shown in the   Appendix:
 \begin{eqnarray}\label{eq:vx1}
 v_x^{A}&=&\sum_{\ell=1}^2\left(E[I(X_{\ell i}>Y_{\ell j})I(X_{\ell i}>Y_{\ell l})]-[E(I(X_{\ell i}>Y_{\ell j}))]^2\right)\nonumber\\
&&\!\!-2\!\left(E[I(X_{1i}\!>\!Y_{1j})I(X_{2i}\!>\!Y_{2l})]\!-\!E[I(X_{1i}\!>\!Y_{1j})]E[I(X_{2i}\!>\!Y_{2l})]\right),\nonumber
 \end{eqnarray}
 and
 \begin{eqnarray}\label{eq:vy1}
 v_y^{A}&=&\sum_{\ell=1}^2\left(E[I(X_{\ell i}>Y_{\ell j})I(X_{\ell k}>Y_{\ell j})]-[E(I(X_{\ell i}>Y_{\ell j}))]^2\right)\nonumber\\
 \!\!&&\!\!-2\!\left(\!E[I(X_{1i}\!>\!Y_{1j})I(X_{2k}\!>\!Y_{2j})]\!-\!E[I(X_{1i}\!>\!Y_{1j})]E[I(X_{2k}\!>\!Y_{2j})]\right).\nonumber
 \end{eqnarray}
%
%
 The optimal ratio for evaluating one marker, say marker 1,  is simply $$\sqrt{\frac{E[I(X_{1i}>Y_{1j})I(X_{1i}>Y_{1l})]-[E(I(X_{1i}>Y_{1j}))]^2}{E[I(X_{1i}>Y_{1j})I(X_{1k}>Y_{1j})]-[E(I(X_{1i}>Y_{1j}))]^2}}.$$ \cite{Janes:06} derive this ratio  in terms of placement values  as
  $$ \sqrt{\frac{Var(S_{\bar{d},1}(Y_{1j}))}{Var(S_{d,1}(X_{1i}))}}.$$

 When $W(u)=I\{u=u_{0}\}$, where $0<u_{0}<1$,  the  $\hat\Delta$-statistic compares the sensitivities at the FPR $u_0$. The optimal ratio in (\ref{eq:rwauc}) reduces to
  $$r_{s}^{*}=\sqrt{\frac{\sum_{\ell =1}^2\left\{R_\ell(u_{0})\}-[R_\ell(u_{0})\}]^{2}\right\}-2A}
  {\sum_{\ell=1}^2\left\{R_\ell'(u_0)^2-[R_\ell'(u_0)u_{0}]^2\right\}-2B}},$$
  where $$A=Pr(X_{1i}>G_1^{-1}(u_{0}),X_{2i}>G_2^{-1}(u_{0}))-R_1(u_{0})R_2(u_{0})$$ and $$B=R_1'(u_0)R_2'(u_0)[Pr(X_{1i}>G_1^{-1}(u_{0}),X_{2i}>G_2^{-1}(u_{0}))-u_{0}^2].$$ The optimal ratio for evaluating marker 1 at the FPR $u_0$ is reduced to the ratio derived in \cite{Janes:06} as $  \sqrt{ R_1(u_0)(1-R_1(u_0))/[u_0(1-u_0)]}/R_1'(u_0)$.
\subsection{Optimal  sampling  ratio for comparing two ordinal markers}\label{sec:ratio2}
The variance of the $\hat \Delta$-statistic involves the first derivatives of the ROC curves. The optimal ratio in (\ref{eq:rwauc}) cannot be readily applied to the ordinal data which often occur  in radiology. In addition, the $\hat\Delta$-statistic does not allow for ties in marker observations.  We thus consider the nonparametric statistic   by \cite{Delong:88} to obtain the optimal ratio for comparing two ordinal markers which are usually two imaging modalities in radiology.   DeLong's statistic estimates $P(X_{1i}>Y_{1j})-P(X_{2i}>Y_{2j})+ [P(X_{1i}=Y_{1j})-P(X_{2i}=Y_{2j})]/2$, and is given as:
\begin{equation}
\hat{\Delta}^{D}
=\frac{1}{mn}\sum^{n}_{j=1}\sum^{m}_{i=1}[\psi(X_{1i},Y_{1j})-\psi(X_{2i},Y_{2j})]\nonumber,
\end{equation}
where  $\psi(X_{\ell i},Y_{\ell j})=1,$ for $Y_{\ell j}<X_{\ell i}$; 1/2 for $Y_{\ell j}=X_{\ell i}$; and 0 for $Y_{\ell j}>X_{\ell i}$, for marker $\ell, \ell=1,2.$
 Let $\Omega^A_\ell$ be $P(X_{\ell i}>Y_{\ell j}) +  P(X_{\ell i}=Y_{\ell j})/2$ for marker $\ell$, and $\hat \Omega^A_\ell$ be its estimator.
\cite{Delong:88} show that the large sample variance of $\hat{\Delta}^{D}$ has the form of
$var(\hat{\Delta}^{D})=v_{x}^{D}/m + v_{y}^{D}/n$, with
\begin{align*}
v_x^{D}=&\frac{1}{m-1}\sum^{m}_{i=1}\Big\{ [\frac{1}{n}\sum_{j=1}^{n}\psi(X_{1i},Y_{1j})-\widehat{\Omega}^A_1 ]^2+ [\frac{1}{n}\sum_{j=1}^{n}\psi(X_{2i},Y_{2j})-\widehat{\Omega}^A_2 ]^2\\
   & \qquad -2 [\frac{1}{n}\sum_{j=1}^{n}\psi(X_{1i},Y_{1j})-\widehat{\Omega}^A_1 ] [\frac{1}{n}\sum_{j=1}^{n}\psi(X_{2i},Y_{2j})-\widehat{\Omega}^A_2 ]\Big\},
\end{align*}
and \begin{align*}
v_y^{D}=&\frac{1}{n-1}\sum^{n}_{j=1}\Big\{ [\frac{1}{m}\sum_{i=1}^{m}\psi(X_{1i},Y_{1j})-\widehat{\Omega}^A_1 ]^2+ [\frac{1}{m}\sum_{i=1}^{m}\psi(X_{2i},Y_{2j})-\widehat{\Omega}^A_2 ]^2\\
  & \qquad
  -2 [\frac{1}{m}\sum_{i=1}^{m}\psi(X_{1i},Y_{1j})-\widehat{\Omega}^A_1 ] [\frac{1}{m}\sum_{i=1}^{m}\psi(X_{2i},Y_{2j})-\widehat{\Omega}^A_2 ]\Big\}.
\end{align*}
Therefore, it follows from (\ref{eq:opti}) that the   ratio, $ r_D^{* }=\sqrt{ v_x^{D}/v_y^{D}},$   maximizes  the power for comparing two ordinal markers.
\section{A two-stage procedure to obtain the optimal ratio}\label{sec:2stage}
\begin{figure}[!h]
\begin{center}
 \includegraphics[scale=0.7]{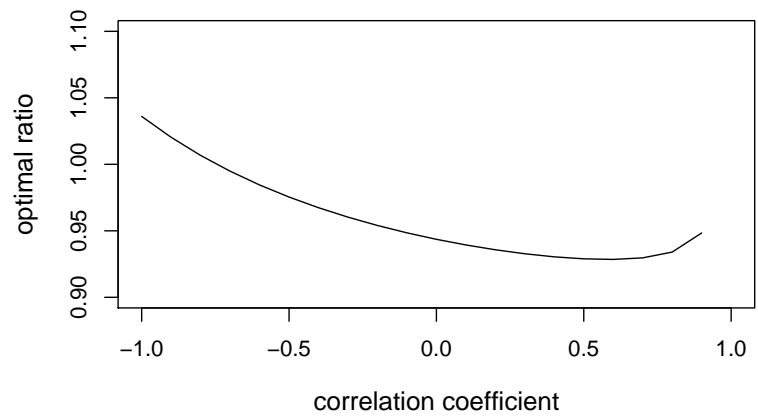}\\
 \caption{Optimal sampling ratio for comparing pAUCs. The observations are from two bivariate normal distributions. The  FPR is  between 0 and 0.6.}
 \label{fig:ratios}
  \end{center}
\end{figure}

One may assume  a parametric model to obtain the variances and resulting optimal ratios derived in the preceding section. When a parametric
model is correctly specified, the optimal ratio  can be calculated
from (\ref{eq:opti}) for comparing ROC summary measures, and the
sample size to obtain a specified power  can be subsequently
derived. However, if the parametric model is mis-specified, the
calculated sample size may not give the appropriate power.
  We calculated  the optimal ratios for comparing the AUCs or pAUCs  from binormal and bi-exponential  distributions. When comparing the AUCs, the
optimal ratio is  close to 1 for a wide range of  the
correlation parameter values for bivariate normal distributions. This implies that equal sampling for two groups  yields the maximum power for a fixed total required sample size. However, the optimal ratio is around 1.5 for bi-exponential distributions, indicating that   sampling 50\% more in  cases than controls
  yields the maximum power to detect
a  difference between markers. When comparing the pAUCs,
Figure
\ref{fig:ratios} shows  the optimal ratios for bivariate
normal distributions.  The optimal sampling ratio varies from $0.94$ to  $1.03$ when correlation coefficients
between two markers vary from $-1$ to $1$. Based on these two examples,    the mis-specification of parametric models at the planning
stage may lead to an incorrect  optimal ratio.


\cite{proschan:04} introduces the concept of internal pilot data which often refers to  accumulated data after a trial is carried out for a certain period of time. To correct for the model mis-specification at the beginning of the trial, we propose a two-stage procedure to use internal pilot data after some observations are available during the trial.
Suppose the total required sample size $N$ is fixed.  Without loss of generality, we use a two-sided test in the proposed procedure. The  procedure
is given in the following steps:
 \begin{itemize}
\item {\it Step 1}: Specify a parametric model to obtain  $v_{x,0}$ and $v_{y,0}$,  and  the resulting initial optimal ratio, $r^*_0 =\sqrt{v_{x,0}/v_{y,0}}$.
\item {\it Step 2}: Use the ratio $r^*_0$ together with $v_{x,0}$, $v_{y,0}$    in the following sample size formula  to calculate initial sample sizes $m_0$ and $n_0$   with power $1-\beta$ and the significance level $\alpha$: \begin{eqnarray}\label{eq:size}
m_0
 =
  \frac{(z_{ \alpha/2}
 +z_{\beta}) ^2 (v_x+r^*_0v_y) }
 {  \Delta_1 ^2},
\end{eqnarray} and $n_0=N-m_0$, where $\Delta_1$ is the
difference between ROC summary measures  under the
alternative hypothesis.
\item {\it Step 3}:  After sufficient marker measurements are
available on $m_1$ cases and $n_1$ controls  at the first stage, the variance expressions
of either the $\Delta$-statistic \citep{wgjj:89} or DeLong's
statistic \citep{Delong:88} are re-calculated using available data. These variance estimators, $\hat{v}_{x,1}$ and $\hat{v}_{y,1} $,  are applied in (\ref{eq:opti}) to
re-calculate the optimal ratio, $\hat{r}^*=\sqrt{ \hat{v}_{x,1}/\hat{v}_{y,1}}.$
\item {\it Step 4}: Continue the trial by recruiting $M_2$ cases and $N_2$ controls, where $M_2$ and $N_2$ are given by
\begin{equation}\label{eq:newsize}
M_2=\frac{N\hat{r}^*}{1+\hat{r}^*}-m_1\quad\textrm{and} \quad
N_2=\frac{N}{1+\hat{r}^*}-n_1.
\end{equation}
\end{itemize}

It is showed in \cite{proschan:04}  that using the internal pilot data for comparing population means in clinical trials maintains the nominal type I error rate. The reason is that the sample variance obtained at the end of the first stage does not give any information for the sample mean  at the end of the trial. The same relationship between the estimated variance and the test statistic is also true for   the $\Delta$-statistic or  DeLong's statistic, as stated in  Proposition 1. The proof is provided in the Appendix.

\noindent\textit{Proposition 1: At the first stage when
$m_1$ and $n_1$  get large, the variance estimated at   the
first stage does not give any information for the  $\Delta$-statistic or  DeLong's statistic  at
the end.}

Proposition 1 shows that estimating variances and the resulting optimal ratio using data from the first stage do  not reveal information about the estimated difference between two ROC statistics obtained at the end of the second stage. Thus, although the optimal ratio is updated during the trial, the analysis at the end of the trial can be carried out in the same fashion as in the trial without updating the optimal ratio. This is important in  maintaining the proper type I error rate.

\section{Example}\label{sec:ex}
In this section, we applied our method to a cancer diagnostic trial  \citep{goddard:90}.  In this study 135 cancer patients and 218 non-cancer patients were recruited. A traditional biomarker, A,  and  newly developed diagnostic biomarkers  were used to test blood samples from each subject. The unit of measurement was mmol of product per minute per millilitre, IU/mm.   Measurements are highly skewed for all the methods. We compared  a new biomarker D and the reference biomarker A to illustrate the power increment and the sample size savings by using the proposed procedure.  We assumed a
contrast of $\Delta_1=0.05$ between AUCs and the type I error rate 0.05  for power and sample size calculation based on a two-sided alternative. At the first stage, we
accrued data on  $m_1=60$ cancer and $n_1=60$ noncancer patients, and
obtained the variance estimates,  $\hat{v}_{x,1}=0.082$ and $\hat{v}_{y,1}=0.035$,
which resulted in the optimal case-control ratio, $\hat{r}^*=1.53$, from
 (\ref{eq:opti}). Let $N$ be the overall sample size, which is 353 by summing the numbers of cases and controls.
 {Using this optimal ratio in the expression   (\ref{eq:newsize}) in Step 4 of the proposed procedure,  the numbers of the cases and controls to be recruited in the second stage were calculated to be  153 and 80, respectively. }
 The power using the optimal ratio was then     50.9\%  using the following equation:
$$1-\beta=\Phi\left( \Delta_1 \sqrt{\frac{N\hat{r}^*}{(1+\hat{r}^*)(\hat{v}_{x,1}+\hat{v}_{y,1}\hat{r}^*)}} -z_{\alpha/2}\right).$$
This power offers 7\% increment over the  power   43.8\%  calculated    using the equation above by replacing $\hat r^*$ with the original  case-control ratio of 0.62. We also investigated the savings on the overall sample size by using the proposed procedure. Using   the original power  43.8\%    with the  estimated optimal ratio, $\hat r^*=1.53$, the overall sample size was calculated to be   to  292 with 177 cancer patients and 115 noncancer patients. This offers  savings of 61 patients over the original ratio.

\section{Simulation studies}\label{sec:sim}

 In this section, we demonstrated the performance of our method for maximizing power when comparing summary statistics of diagnostic tests.
  We compared the proposed  two-step procedure with the equal case-control ratio and a fixed case-control ratio under three parametric models.
  Three pairs of AUCs and pAUCs were specified in advance. We used  DeLong's statistic   for comparing the AUCs  and the $\Delta$-statistic  for comparing the pAUCs.
We simulated 5000 observations  from  bivariate normal (BN),
bivariate lognormal (LN) and bivariate exponential (BE)
distributions, respectively. The bivariate normal models had the
forms of $(X_1,X_2)\sim N\{(\mu_1,\mu_2),\Sigma\}$ and
$(Y_1,Y_2)\sim N\{(0,0),\Sigma\}$, where in the $2\times 2$ matrix $\Sigma$, the diagonal elements  are 1's and off-diagonal elements are $\rho$.
We chose $\rho=0.1$ and $\rho=0.25$ in our simulations. $\mu_1$ and
$\mu_2$ were computed according to three pairs of AUCs, $(0.70,0.75)$, $(0.75,0.80)$ and $(0.70,0.80)$, respectively. For comparing the pAUCs with the FPR in the range of $(0,0.6)$,  $(\mu_1,\mu_2)$
were used for three pairs of pAUCs,  $(0.30,0.35)$, $(0.35,0.40)$ and $(0.30,0.40)$, respectively.
    The bivariate lognormal models had the forms of
$exp(X_1,X_2)$ and $exp(Y_1,Y_2)$ for cases and controls,
respectively. They had the same values of $(\mu_1,\mu_2)$ for the
AUCs and pAUCs as above. And then, according to the algorithm in
Gumbel (1960), the bivariate exponential random variables take the
form $H(x,y)=H_1(x)H_2(y)[1+4\rho\{1-H_1(x)\}\{1-H_2(y)\}]$, where
$\rho\in [-0.25,0.25]$. We set $\rho$ be $0.1$ and $0.25$ here. The
marginal survival functions for cases and controls were
$exp(-\beta_{\ell 1}x)$ and $exp(-\beta_{\ell 2}y)$, so we could
generate data from these two distributions respectively. In the
simulation, we set $\beta_{1 1}=1$ and $\beta_{2 1}=1 $.
$\beta_{12}$ and $\beta_{22}$ were computed according to the AUC or
pAUC values. For the pairs of AUCs $(0.70,0.75)$, $(0.75,0.80)$, and
$(0.70,0.80)$, the corresponding $(\beta_{12},\beta_{22})$ values were
$( 2.333, 3.003),(3.003, 4.000)$ and $(2.333, 4.000)$. For the pairs
of pAUCs $(0.30,0.35)$, $(0.35,0.40)$ and $(0.30,0.40)$, the
$(\beta_{12},\beta_{22})$ values were $(1.8957, 2.5094)$,
$(2.5094,3.3887)$ and $(1.8957, 3.3887)$, respectively.

In our simulation, we first assumed that our samples were from
bivariate normal distributions, then used equation (\ref{eq:size})
to calculate the initial total required sample size. With the type I
error rate 0.05 and power  $80\%$, the initial
total required sample sizes were $N= 1421, 1200, $ or $326$ to detect the difference
of three pairs of AUCs of $(0.70,0.75)$, $(0.75,0.80)$ and $(0.70,0.80)$, respectively,
with $\rho=0.1$. When $\rho=0.25$, the total required sample sizes,
$N=1207, 1025,$ or  278, were needed to detect the difference in these   pairs. For comparing  the pAUCs, the initial total required sample sizes were
$N=1067, 979,$ and  251, for  for $\rho=0.1$, and $N=915, 842$ and  216 for
$\rho=0.25$. There were three different sampling
ratios: 1) the proposed two-stage optimal ratio; 2) fixed sample ratio of 0.5; 3)
equal sampling ratio.  To implement the proposed
method, we defined the number of available observations at the first stage, $m_1=n_1=N /4$. By substituting nonparametric variance
estimates $\hat{v}_{x,1}$ and $\hat{v}_{y,1}$, the resulting optimal ratio
was estimated by $\hat{r}^*= \sqrt{ \hat{v}_{x,1}/\hat{v}_{y,1}},$ and $M_2$ and $N_2$ were calculated using (\ref{eq:newsize}).
We then generated  $M_{2}$ new  observations for cases and $N_{2}$
observations for controls. Subsequently, the null hypothesis of
equal AUCs or pAUCs was rejected  in favor of the alternative if  the $Z$-statistic calculated using
all simulated data  was greater than or equal to
$z_{0.025}$. The simulated power was then calculated as the percent of  times out of 5000 that the null hypothesis was rejected. The simulated powers  for all simulation settings are present in Table
\ref{tb:t1}.

   Table \ref{tb:t1}   illustrates that larger correlations resulted in
higher rejection rates. The sampling
ratio is another factor impacting the power when the alternative
hypothesis is true.   For
different underlying distributions, the proposed two-stage
method has   higher powers  than the fixed ratios in most of the settings.

We also evaluated the performance of the two-step procedure to see whether the procedure maintains the nominal  type I error rate. We used the total required sample sizes,
$N=200$, 400, or 500. The parametric distributions and three
different sampling ratios used in the previous simulation were considered. We assumed equal AUCs or pAUCs with the AUCs being $(0.70, 0.75, 0.80)$, and the pAUCs being $(0.30, 0.35, 0.40)$. The nominal type I error rate was
 $0.05$ in our simulation. The simulated type I error rates are shown in
Table \ref{tb:t4}. All these  rates are close to the nominal level   when the sample size goes to 500.

Variability in the estimators, $\hat{v}_{x,1}$
and $\hat{v}_{y,1}$, is associated with the  initial sample sizes at the first stage. Such variability  affects the calculation of the optimal sampling ratio, which may in turn have an impact on
the power in the proposed procedure. We conducted another simulation study to investigate the impact of  the initial sample size selection. We used the total required sample size of $400$, and set the
initial sample sizes of cases and controls to be $m_0=n_0=50, 60, 80,$ or  100. Observations were simulated from    the binormal
distributions with the difference of $0.05$  between two AUCs. In each simulation, the variance estimators for calculating the optimal ratio  were estimated at the first stage from three scenarios, namely, 1) a single set of $m_0$ cases and $n_0$ controls, 2) averaging variance estimates of 10 sets of $m_0$ cases and $n_0$ controls, and 3) averaging variance estimates of 100 sets of $m_0$ cases and $n_0$ controls.   Results based on 1000 replications for each setting are listed in
Table \ref{tb:t6}. It indicates   some variations in  power for the first scenario.  When more  datasets are involved
in the calculation, power becomes more stable regardless of the initial
sample sizes. More importantly, Table \ref{tb:t6} shows that the initial sample size selection had little impact on the final power.

\section{Conclusion}\label{sec:dis}
 The optimal sampling ratio in diagnostic trials  can maximize  the test power or minimize the overall sample size.  The optimal sampling ratio discussed in this paper is analogous to the optimal allocation ratio in assigning patient treatments in clinical trials. The optimal allocation ratio has been used in clinical trials for decades, but the importance of the optimal ratio in diagnostic trials has not been widely recognized. Implementation requires the calculation of  complicated variances   of frequently used ROC statistics. This paper  discusses a common variance structure for   ROC statistics, and thereby introduces optimal sampling ratios  in  comparative diagnostic trials based on these statistics.  Two popular nonparametric ROC statistics are used to illustrate the explicit forms of the optimal ratios because their variance expressions can be written as the sum of separate terms; one relates to the cases, and the other relates to the controls. The same variance structure is shared by many existing parametric and semiparametric ROC statistics. This implies that the optimal ratio form derived in (\ref{eq:opti}) is also applicable to these existing statistics.

 When marker results follow normal distributions, the optimal  sampling ratio is close to 1 for many parameter settings. Then sampling the same number of cases and controls can potentially achieve the maximal power for a fixed total required sample size. When the marker results follow exponential distributions, the sampling ratio is close to 1.5. We need to sample more cases than controls to gain power or reduce the overall sample size. If preliminary studies are available before carrying out a comparative diagnostic trial, the variance can be estimated using   pilot data to obtain the optimal ratio for comparing specified ROC summary measures. The ratio can then be used to recruit patients in the trial, and re-calculating the ratio may not be necessary during the trial.  However, when   medical practitioners do not have  preliminary data  for the markers and are not certain about the distributions of the marker results, the distribution assumption used for obtaining the optimal ratio may be far from the true underlying distributions for the marker results. This may result in    less power or  larger overall sample sizes than using the true optimal ratio. The proposed two-stage procedure is then particularly useful to ensure that the optimal ratio can be re-calculated using  using internal pilot data during the trial.   The proposed procedure performed well in a large scale simulation study.  We also demonstrated that the proposed procedure maintains the nominal type I error rate in the simulation. We  used an example in cancer diagnostic studies to illustrate the application of our method on maximizing the test power and saving overall sample sizes.  The results indicated that compared with the original sampling ratio, using the   proposed two-stage procedure for a fixed overall sample size increased the test power. Alternatively,   for the fixed test power, the proposed procedure   reduced  the overall sample size by nearly 25\%.

It is sometimes  desired to
  minimize the total cost  in a  diagnostic trial with a limited budget. High cost may be associated with diagnostic trials considering using a gold standard test to identify the subjects and using markers to diagnosing them. This is particularly true in medical imaging diagnostic trials when expensive medical imaging devices costing hundreds of dollars for a single session of scans are involved.  A case may cost more than a control because of  higher expenses associated with providing necessary medical care when classifying and diagnosing them.
We may consider $ c_1$ and $ c_{2}$ as costs related to a case and a control, respectively.  Usually, $c_1$ and $c_2$   can be determined by medical experts before conducting a trial.  Then similar to the derivation  in Section \ref{sec:ratio}, the optimal sampling ratio for minimizing the total cost   is given by
$
 r_{c}^*=\sqrt{ c_2v_x/c_1v_y } $ for a fixed power. This ratio reduces to the one derived in (\ref{eq:opti}) when $ c_1=c_{2}$. An interesting future research topic is to investigate the optimal ratio when the costs are related to the true AUC parameters.

\bibliographystyle{biorefs}
\bibliography{cite_OA}
\section*{Appendix}
\renewcommand \theequation{A.\arabic{equation}}
\setcounter{equation}{0}
{\centering\emph{Appendix:  variance derivation and proof of Proposition 1}}
\par \emph{Derivation of $v_x^A$ and $v_y^A$}

We  can show that
  $$ \int_0^1\!\!\!\int_0^1[S_{d}\{S_{\bar{d}, 1}^{-1}(s), S_{\bar{d}, 2}^{-1}(t)\}]dsdt$$
   can be expressed as
  $$\int_{-\infty}^{\infty}\int_{-\infty}^{\infty} S_{d}(y_1, y_2)dS_{\bar{d}, 1}(y_1)dS_{\bar{d}, 2}(y_2).$$
   Let $S_{\bar{d}, 1}^{-1}(s)=y_1$ and $S_{\bar{d}, 2}^{-1}(t)=y_2$, then, we have
  $$ \int_0^1\!\!\!\int_0^1[S_{d}\{S_{\bar{d}, 1}^{-1}(s), S_{\bar{d}, 2}^{-1}(t)\}]dsdt=E[I(X_{1i}>Y_{1j})I(X_{2i}>Y_{2l})].$$
Similarly, $v_y$   becomes
\begin{eqnarray}
v_y &=&  \sum_{\ell =1}^2 \left[\int_0^1\!\!\!\int_0^1
 r_\ell(s)r_\ell(t) (s\wedge
t)dsdt -\left \{\int_0^1r_\ell(s)s ds \right \}^2\right]\nonumber\\
&& -2 \int_0^1\!\!\!\int_0^1
 r_1(s)r_2(t)[S_{\bar{d}}\{S_{\bar{d}, 1}^{-1}(s),S_{\bar{d}, 2}^{-1}(t)
\}-st ]dsdt\nonumber.
\end{eqnarray}
It follows that
\begin{eqnarray}
&\int_0^1\!\!\!\int_0^1
 r_1(s)r_2(t)S_{\bar{d}}\{S_{\bar{d}, 1}^{-1}(s), S_{\bar{d}, 2}^{-1}(t)\}dsdt\nonumber\\
 =&\int_0^1\int_0^{1}\frac{S_{d, 1}'\{ S_{\bar{d}, 1}^{-1}(s)\}}{S_{\bar{d}, 1}'\{ S_{\bar{d}, 1}^{-1}(s)\}}\frac{S_{d, 2}'\{ S_{\bar{d}, 2}^{-1}(t)\}}{S_{\bar{d}, 2}'\{S_{\bar{d}, 2}^{-1}(t)\}}S_{\bar{d}}\{S_{\bar{d}, 1}^{-1}(s), S_{\bar{d}, 2}^{-1}(t)\}dsdt.\nonumber
 \end{eqnarray}
 Let $S_{\bar{d}, 1}^{-1}(s)=y_1$ and $S_{\bar{d}, 2}^{-1}(t)=y_2$, then it follows that
 \begin{align}
  &\int\int S_{d, 1}'\{ y_1\}S_{d, 2}'\{ y_2\}S_{\bar{d}}(y_1, y_2))dy_1dy_2\nonumber\\
 =&E[I(X_{1i}<Y_{1j})I(X_{2k}<Y_{2j})]\nonumber\\
  =&E[(1-I(X_{1i}>Y_{1j}))(1-I(X_{2k}>Y_{2j}))]\nonumber\\
 =&1-E(I(X_{1i}>Y_{1j}))-E(I(X_{2k}>Y_{2j}))+E[I(X_{1i}>Y_{1j})I(X_{2k}>Y_{2j})]\nonumber.
  \end{align}
Because
 $$\int_0^1\!\!\!\int_0^1r_1(s)r_2(t)stdsdt$$
  can also be written as
 $$1-Pr(X_{1i}>Y_{1j})-Pr(X_{2k}>Y_{2j})+E[I(X_{1i}>Y_{1j})]E[I(X_{2k}>Y_{2j})],$$
 the expressions for  $v_x$ and $v_y$  are simplified as follows:
 \begin{eqnarray}\label{eq:vx1}
 v_x^{A}&=&\sum_{\ell=1}^2\left(E[I(X_{\ell i}>Y_{\ell j})I(X_{\ell i}>Y_{\ell l})]-[E(I(X_{\ell i}>Y_{\ell j}))]^2\right)\nonumber\\
&&\!\!-2\!\left(E[I(X_{1i}\!>\!Y_{1j})I(X_{2i}\!>\!Y_{2l})]\!-\!E[I(X_{1i}\!>\!Y_{1j})]E[I(X_{2i}\!>\!Y_{2l})]\right),
 \end{eqnarray}
 and
 \begin{eqnarray}\label{eq:vy1}
 v_y^{A}&=&\sum_{\ell=1}^2\left(E[I(X_{\ell i}>Y_{\ell j})I(X_{\ell k}>Y_{\ell j})]-[E(I(X_{\ell i}>Y_{\ell j}))]^2\right)\nonumber\\
 \!\!&&\!\!-2\!\left(\!E[I(X_{1i}\!>\!Y_{1j})I(X_{2k}\!>\!Y_{2j})]\!-\!E[I(X_{1i}\!>\!Y_{1j})]E[I(X_{2k}\!>\!Y_{2j})]\right).
 \end{eqnarray}  \hfill $\Box$

\par \emph{Proof of Proposition 1}

 We first prove that the proposition is true for the $\Delta$-statistic. Similar arguments can then be used for the Delong's statistic. Let $\bar w=\frac{\sum^{m_{1}}_{i=1}w_{i}}{m_{1}}$.
  We   see that in (4), $w_{i}$'s are $i.i.d.$ random variables independent of $i.i.d.$ random variables $v_{j}$'s. It then follows that
 \begin{eqnarray}
 cov\!\left(\frac{\sum^{m_{1}+m_{2}}_{i=1}w_{i}}{m_{1}+m_{2}},  w_{i}\!-\!\bar w\!\right)\!=\!
cov\!\left(\frac{\sum^{m_{1}+m_{2}}_{i=1}w_{i}}{m_{1}+m_{2}},
w_{i}\!\right)\!-\!
cov\!\left(\frac{\sum^{m_{1}+m_{2}}_{i=1}w_{i}}{m_{1}+m_{2}},
\bar w\!\right)\nonumber
\end{eqnarray}
equals $0$, and
\begin{align*}
&\bar{w}=\int_{0}^{1}S_{d,1}(S_{\bar{d},
1}^{-1}(u))du-\frac{1}{m_{1}}\sum_{i=1}^{m_{1}}\int_{0}^{1}I(X_{1i}\leq
S_{\bar{d}, 1}^{-1}(u))du\\
&  +\frac{1}{m_{1}}\sum_{i=1}^{m_{1}}\int_{0}^{1}I(X_{2i}\leq
S_{\bar{d}, 2}^{-1}(u))du-\int_{0}^{1}S_{d,2}(S_{\bar{d}, 2}^{-1}(u))du.\\
\end{align*}
We then   get
 \begin{align*}
&(w_{i}-\bar{w})\!=\!\frac{1}{m_{1}}\sum_{i=1}^{m_{1}}\int_{0}^{1}I(X_{1i}\leq
S_{\bar{d}, 1}^{-1}(u))du-\int_{0}^{1}I(X_{1i}\leq
S_{\bar{d}, 1}^{-1}(u))du\\
&  \qquad+\int_{0}^{1}I(X_{2i}\leq S_{\bar{d},
2}^{-1}(u))du-\frac{1}{m_{1}}\sum_{i=1}^{m_{1}}\int_{0}^{1}I(X_{2i}\leq
S_{\bar{d}, 2}^{-1}(u))du.
\end{align*}
Therefore,
\begin{align*}
&\frac{\sum_{i=1}^{m_{1}}(w_{i}-\bar{w})^{2}}{m_{1}-1}
 \approx\hat{v}_{x_{1}}/m_{1},
\end{align*}
which indicates that for large sample sizes, $(\sum^{m_{1}+m_{2}}_{i=1}w_{i})/(m_{1}+m_{2})$
is independent of $\hat{v}_{x,1}/m_{1}$. Similarly, we get that for large sample sizes,
$(\sum^{n_{1}+n_{2}}_{j=1}v_{j})/(n_{1}+n_{2})$  is
independent of $\hat{v}_{y,1}/n_{1}$.
  \hfill $\Box$

\begin{table}[!h]
\caption{Simulated power  (in \%)   for
comparing AUCs or pAUCs}\label{tb:t1}
\begin{tabular}{|c|c c| c c c c| c c |c c |}
\hline \multicolumn{11}{c}{Comparing AUCs using the DeLong's method}\\\hline
\hline \multicolumn{1}{|c|}{}                        &
 &&\multicolumn{4}{c|}{Two-Stage }&\multicolumn{2}{ c|}{$r=1$}&\multicolumn{2}{ c|}{$r=0.5$}\\
 \cline{2-11}
 \multicolumn{1}{|c|}{$\rho$}                        &
 $ $  &$\Omega^{A}_{1}\setminus\Omega^{A}_{2}$ & 0.70  & $AR$
  & 0.75   & $AR$ & 0.70   & 0.75  & 0.70  & 0.75   \\
\hline \multicolumn{1}{|c|}{}                        & $BN$ & 0.75 &
79.3& 1.001& -  & -
            &79.7 & - & 74.6  & -\\

 $$  &  &0.80   & 81.0 & 1.003  &79.1  & 1.002
             & 78.8 & 80.5  &75.3  &74.8\\

\multicolumn{1}{|c|}{0.1}                        & $LN$ & 0.75 &
80.5 & 1.001& -  & -
            &80.2 & - & 75.6  & -\\

 $$  &  &0.80   & 80.3 & 1.003  &79.4  &1.000
            & 79.8 &80.6  &75.3  &75.2\\

\multicolumn{1}{|c|}{}                        & $BE$ & 0.75 & 81.0 &
1.340& -  & -
            &80.4 & - & 71.2 & -\\

  $$ &  &0.80   & 81.6 & 1.467  &81.8  & 1.551
            & 80.0 & 80.4 &70.0  &69.9\\

\hline \multicolumn{1}{|c|}{}                        &
 $ $  &$\Omega^{A}_{1}\setminus\Omega^{A}_{2}$ & 0.70  & $AR$
  & 0.75   & $AR$ & 0.70  & 0.75   & 0.70   & 0.75  \\
 \cline{2-11}
\multicolumn{1}{|c|}{}                        & $BN$ & 0.75 & 79.8 &
1.002& -  & -
            &80.0 & - & 74.9 & -\\

 $$  &  &0.80   & 80.2 & 1.007 &80.3  &1.004
             & 79.1 & 80.2 &75.1  &74.5\\
\multicolumn{1}{|c|}{0.25}                        &

$LN$ & 0.75 &79.8 & 1.002& -  & -
            &79.8 & - & 75.3  & -\\

 $$  &  &0.80   & 80.2 & 1.005 &79.8 &1.003
             & 79.7 &79.6  &74.9  &75.3\\
\multicolumn{1}{|c|}{}                        &

$BE$ & 0.75 &83.7 &1.412& -  & -
            &82.6 & - & 74.2  & -\\

 $$ &   &0.80   &83.6 & 1.482 &83.5 &1.579
             &82.8 & 81.0 &72.5  &71.4\\

\hline \multicolumn{11}{c}{Comparing pAUCs using the $\Delta$-statistic}\\\hline
\hline \multicolumn{1}{|c|}{}                        &
 &&\multicolumn{4}{c|}{Two-Stage}&\multicolumn{2}{ c|}{$r=1$}&\multicolumn{2}{ c|}{$r=0.5$}\\
 \cline{2-11}
\multicolumn{1}{|c|}{$\rho$}                        &
 $ $  &$\Omega^{PA}_{1}\setminus\Omega^{PA}_{2}$ & 0.30   & $AR$
  & 0.35  & $AR$ & 0.30   & 0.35   & 0.30   & 0.35   \\
\hline \multicolumn{1}{|c|}{}                        & $BN$ & 0.35 &
79.2 & 0.952& -  & -
            &78.7 & - & 73.9  & -\\

 $$ &   &0.40   & 79.8 & 1.008  &79.3  & 0.954
             & 80.6 & 79.2  &75.1  &75.0\\

\multicolumn{1}{|c|}{0.1}   & $LN$ & 0.35 & 79.0 & 0.953& -  & -
            &78.9 & - & 74.8  & -\\

  $$  & &0.40   & 80.4  & 1.003  &79.3  &0.954
            & 81.0 &80.6  &75.8  &74.8\\

\multicolumn{1}{|c|}{}                        & $BE$ & 0.35 & 84.6 &
1.249& -  & -
            &84.0 & - & 76.6 & -\\

$$  &   &0.40   & 84.9 & 1.389  &83.6  & 1.386
            & 84.6 & 83.9 &76.6  &74.4  \\

\hline \multicolumn{1}{|c|}{}   &
 $ $  &$\Omega^{PA}_{1}\setminus\Omega^{PA}_{2}$ & 0.30   & $AR$
  & 0.35   & $AR$ & 0.30   & 0.35   & 0.30   & 0.35   \\
 \cline{2-11}
\multicolumn{1}{|c|}{}   & $BN$ & 0.35 & 78.7 & 0.947& -  & -
            &78.0 & - & 75.4  & -\\

 $$   &  &0.40   & 80.6 & 1.014 &78.7 &0.952
             & 80.5 & 79.8 &76.0  &75.2 \\

\multicolumn{1}{|c|}{0.25}   & $LN$ & 0.35 &79.2 & 0.949& -  & -
            &79.0 & - & 74.6  & -\\

  $$ &   &0.40   & 80.9 & 1.013 &79.3 &0.950
             & 80.7 &79.2  &76.6  &75.3\\

\multicolumn{1}{|c|}{}   & $BE$ & 0.35 &86.9 &1.219& -  & -
            &87.1 & - & 80.0  & -\\

  $$ &   &0.40   &86.8 & 1.385 &84.5 &1.365
             &86.6 & 84.0 &78.6  &76.9\\

\hline
\end{tabular}

AR - the average ratio, BN - bivariate normal, LN - bivariate
lognormal, BE - bivariate exponential, $\Omega^{A}_{1}$ - the AUC
for marker 1, $\Omega^{A}_{2}$ - the AUC for marker 2, $\Omega^{PA}_{1}$ - the
pAUC for marker 1, $\Omega^{PA}_{2}$ - the pAUC for marker 2, $ \rho$ - the
correlation coefficient of two markers.
\end{table}

\begin{table}[!h]
\caption{Type I error rates (in \%) for comparing the
AUCs or pAUCs}\label{tb:t4}
\begin{tabular}{cccrrrrrrrr}
\hline
      &       & \multicolumn{ 4}{c}{Comparing the AUCs} &       & \multicolumn{ 4}{c}{Comparing the pAUCs} \\
\hline
$\rho$ &       & AUCs  & \multicolumn{1}{|c|}{N=200} & 400   & 500   &       & pAUCs & \multicolumn{1}{|c|}{N=200} & 400   & 500 \\
\hline
      &     BN     & 0.70  & 4.5   & 5.8   & 5.0   &       & 0.30  & 5.6   & 5.9   & 4.6 \\
      &        & 0.75  & 5.4   & 5.8   & 4.0   &       & 0.35  & 6.8   & 5.8   & 5.0 \\
      &         & 0.80  & 5.3   & 5.1   & 5.2   &       & 0.40  & 6.7   & 5.5   & 6.1 \\
      &     LN     & 0.70  & 5.7   & 5.3   & 5.0   &       & 0.30  & 6.8   & 5.9   & 4.6 \\
0.10  &         & 0.75  & 5.4   & 6.5   & 4.3   &       & 0.35  & 6.8   & 5.8   & 5.0 \\
      &         & 0.80  & 5.3   & 4.4   & 5.9   &       & 0.40  & 6.7   & 5.5   & 6.1 \\
      &     BE     & 0.70  & 5.4   & 5.1   & 4.4   &       & 0.30  & 6.2   & 5.4   & 5.5 \\
      &        & 0.75  & 5.3   & 6.6   & 4.0   &       & 0.35  & 7.7   & 7.4   & 5.6 \\
      &        & 0.80  & 5.2   & 5.1   & 4.3   &       & 0.40  & 7.5   & 7.3   & 5.0 \\
\hline
      &     BN     & 0.70  & 4.9   & 4.8   & 4.7   &       & 0.30  & 5.9   & 5.3   & 5.4 \\
      &        & 0.75  & 5.9   & 5.4   & 5.9   &       & 0.35  & 5.2   & 5.3   & 5.2 \\
      &        & 0.80  & 5.5   & 6.1   & 5.6   &       & 0.40  & 5.6   & 5.8   & 5.1 \\
0.25  &     LN     & 0.70  & 5.2   & 5.3   & 5.5   &       & 0.30  & 6.0   & 5.3   & 5.4 \\
      &        & 0.75  & 5.9   & 5.4   & 5.9   &       & 0.35  & 5.2   & 5.3   & 5.2 \\
      &        & 0.80  & 5.8   & 3.9   & 4.3   &       & 0.40  & 6.7   & 5.8   & 5.4 \\
      &     BE     & 0.70  & 4.2   & 5.0   & 3.9   &       & 0.30  & 5.0   & 5.6   & 4.8 \\
      &        & 0.75  & 5.3   & 5.7   & 4.4   &       & 0.35  & 5.7   & 7.0   & 6.4 \\
      &         & 0.80  & 5.2   & 5.1   & 4.3   &       & 0.40  & 6.8   & 6.8   & 6.4 \\
\hline
\end{tabular}\\
BN - bivariate normal,  LN  - bivariate lognormal,
 BE  - bivariate exponential, $N$ - the total required sample size, $ \rho$ - the correlation coefficient of two markers
\end{table}

\begin{table}[!h]
\caption{Power comparison}\label{tb:t6}
\centering\begin{tabular}{|c|c|llll|}
\hline
&&& $m_{0} (n_{0})$&& \\
\hline
$K$&$$&50 & 60 &80 &100  \\
\hline
$1 $ &Ratio &1.02 & 1.01 & 0.87& 1.07  \\
$$&Power (\%) &31.9  &26.6  &32.6  &33.7  \\
\hline
 $10 $&Ratio &1.03 & 0.94& 1.03& 1.05  \\
$$&Power (\%)&29.9  &31.3  &31.9  &32.5  \\
\hline
$100$& Ratio &1.03 & 1.02& 1.00 & 1.03 \\
$$& Power (\%) &31.1  &30.8  &31.1  &31.4  \\
\hline
\end{tabular}\\
$K$ - the number of datasets simulated to estimate variances.
\end{table}

\end{document}